\documentclass[aps,pra,twocolumn,showpacs,nofootinbib,longbibliography,notitlepage]{revtex4-2}
\usepackage{etex}
\usepackage{amsmath,amssymb,amsthm}
\usepackage[colorlinks=true,citecolor=blue,urlcolor=blue]{hyperref}
\usepackage[pdftex]{graphicx}
\usepackage{times,txfonts}
\usepackage{braket}
\usepackage{color}
\usepackage{natbib}
\usepackage{amsmath,blkarray}
\usepackage{mathtools}
\usepackage{latexsym}
\usepackage{tabularx, booktabs}
\usepackage{graphics,epstopdf}
\usepackage{graphicx}
\usepackage{float}
\usepackage{graphicx}
\usepackage{amsfonts}
\usepackage{subcaption}
\usepackage{color,soul}

\newcommand{\na}{\nonumber}

\newcommand{\be}{\begin{equation}}
\newcommand{\ee}{\end{equation}}
\newcommand{\ba}{\begin{eqnarray}}
\newcommand{\ea}{\end{eqnarray}}

\begin{document}
	\title{Self-testing of an unbounded number of mutually 
 commuting {local}  observables }
		\author{Sneha Munshi}
	\author{ A. K. Pan }
	\email{akp@phy.iith.ac.in}
	\affiliation{Indian Institute of Technology Hyderabad, Kandi, Sangareddy, Telangana 502285, India}
	
	\begin{abstract}
       Based on the optimal quantum violation of suitable {Bell's inequality}, the device-independent self-testing of state and observables has been reported. It is well-studied that locally commuting or compatible observables cannot be used to reveal quantum nonlocality.  Therefore, the self-testing of commuting {local}  observables cannot be possible through the Bell test.  In this work, we demonstrate the self-testing of a set of mutually commuting {local} observables. Such certification has not hitherto been reported. We show that the optimal quantum violations of suitably formulated bilocality and $n$-locality  inequalities in networks uniquely fix the observables of one party to be mutually commuting. In particular, we first demonstrate that in a two-input-arbitrary-party star network, two commuting  {local} 
 observables can be self-tested. Further, by considering an arbitrary-input three-party bilocal network scenario, we demonstrate the self-testing of an unbounded number of mutually commuting {local} observables. 
       \end{abstract}
	\maketitle\section{Introduction}

The existence of non-commuting observables is a distinctive feature of quantum theory from its classical counterpart. Such a quantum feature plays a pivotal role in quantum information processing and cryptographic tasks. In quantum theory, the joint probability of two non-commuting observables does not exist, and any measurement of the prior observable influences the measurement of the posterior observable \cite{buschrep}. The incompatibility is a weaker notion of non-commutativity and is mainly motivated by the perspective of quantum measurement. Two non-commuting observables can be compatible, i.e., jointly measurable, depending on the suitably defined weakness of the measurement. However, two commuting observables are compatible as joint probability always exists, and the measurement of one does not disturb the other.

The demonstration of Bell's theorem \cite{bell} requires the relevant local observables to be non-commuting or, more generally, incompatible. The quantum violation of two-input-two-output Clauser-Horne-Shimony-Holt 
 (CHSH) inequality \cite{chsh} in the simplest Bell scenario inevitably requires the local observables to be non-commuting. The optimal quantum violation is achieved when they are anticommuting. It is proved \cite{wolf} that any pair of two-outcome incompatible measurements can violate CHSH inequality. 
 However, for more than two input scenarios, this correspondence breaks down \cite{bene}.  Given an arbitrary set of non-commuting observables, it is not a priori clear whether a suitable Bell's inequality can always be formulated to demonstrate the nonlocality by violating the said inequality.   

 Besides the immense conceptual insights Bell's theorem adds to the quantum foundations research, it provides a multitude of practical applications in quantum information processing (see, for extensive reviews, \cite{hororev,guhnearev,brunnerrev}). Moreover, the {nonlocal correlations are} device-independent, i.e., no characterization of the devices is needed to be assumed. Only the observed output statistics are enough to certify nonlocality. The device-independent {nonlocal correlations} are used as a resource for secure quantum key distribution \cite{bar05,acin06,acin07,pir09}, randomness certification \cite{col06,pir10,nieto,col12}, witnessing Hilbert space dimension \cite{wehner,gallego,ahrens,brunnerprl13,bowler,sik16prl,cong17,pan2020} and for achieving advantages in communication complexity tasks \cite{complx1}.

The maximum quantum value of a given Bell expression enables device-independent certification - commonly known as self-testing \cite{mayer98}.  For a recent review of self-testing, we refer the reader to Ref. \cite{supicrev}.  In its traditional form, self-testing is a device-independent protocol that aims to uniquely characterize the nature of the target quantum state and measurements solely from the input-output correlations. Essentially, this requires finding a suitable Bell's inequality whose maximum violation is achieved uniquely by the target state and measurements involved.  In other words, the state and measurements are device-independently certified with minimal assumptions, i.e., the devices are  uncharacterized (the so-called black-boxes) and the dimension of the system remains unspecified.  Once we obtain the maximum quantum value of a Bell inequality  that eventually guarantees the extremal points in the polytope made by behaviour of the joint probabilities.    For example, the optimal violation of CHSH inequality self-tests the maximally entangled state and mutually anti-commuting local observables.  The self-testing scenario was first proposed by Mayers, and Yao \cite{mayer98}. Later, McKague and Mosca \cite{mckague} used this isometric embedding to develop a generalized Mayers-Yao test \cite{mayers}.    Since then, a flurry of works on this topic has been reported \cite{wagner2020,farkas2019, mckague12,wu,mckague16,and17,cola,supic18,bowels18prl,bowles18,coopmans,tavakoli19a,tavakoli18,farkas,kartik,mir19,tavakoli20,smania,paw20,gomez16,gomez18,Miklin2020,Anwar2020,Foletto2020,sumit21,Abhy2023}. 

Note that, device-independent certification is quite  challenging to implement experimentally. The semi-device-independent prepare-measure protocols with bounded dimensions are constructed which are experimentally less cumbersome. Self-testing quantum states and measurements in the prepare-measure scenario have been demonstrated in \cite{tavakoli18,farkas}. Quite a number of works self-tested the non-projective measurements in device-independent or semi-device-independent scenarios \cite{mir19,tavakoli20,smania,paw20,gomez16,gomez18,pan2021}. Semi-device-independent self-testing of an unsharp instrument through the sequential measurements has also been reported \cite{kartik,Miklin2020,Anwar2020,Foletto2020,sumit21,Abhy2023}. Recently, device-independent certification of the unsharp instrument is also reported \cite{prabuddha}. 

{In the network Bell tests\cite{Branciard10,Branciard12}, nontrivial forms of nonlocal correlations arise that cannot be traced back to the standard multipartite  Bell scenario due to the independence condition of the sources. The set of quantum correlations in a network, in general, becomes non-convex, thereby making the characterization of nonlocality and self-testing more complicated compared to the standard Bell test.  The above issues have been addressed in detail in \cite{Nava2020,Agre2021} and the self-testing argument based on network nonlocality has been discussed. Recently, considering the bilocal network scenario, it has been shown that how genuine nonlocal correlations enable self-testing of quantum state and observables \cite{Supic2021}. The self-testing of all entangled states has also been demonstrated using quantum correlation in network \cite{Supic2023}.}    

The aim of this paper is to certify a set of mutually commuting {local} observables. Such certification has hitherto not been demonstrated. It is a common perception that commuting observables do not provide non-classicality. As discussed above, no violation of Bell's inequality can be obtained if the {local} observables are mutually commuting. Against this backdrop, in this work, we propose self-testing schemes that certify an unbounded set of mutually commuting  {local} observables.  This is done through the optimal quantum violations of suitable network bilocal and $n$-local inequalities which are achieved when one observer performs the measurement of $m$ mutually commuting  {local} observables where both $n$ and $m$ is arbitrary. In other words, we have shown that the optimal quantum violation of certain network inequalities uniquely fix a set of mutually commuting   observables and thus the corresponding set of observables are being self-tested, independent of the dimension.  

We first consider the star network   featuring arbitrary $n$  independent sources, $n$  edge parties, and a central party.  {Each source distributes a physical state with an edge party and the central party and  each party  performs two binary-outcome measurements}. We demonstrate that the optimal quantum violation of a suitable $n$-locality inequality certifies that the two observables of the central party are commuting when $n$ is even. Since our derivation of optimal value is dimension independent, we demonstrate the self-testing of two commuting observables in arbitrary dimensions.  


Further, we propose the self-testing of an arbitrary number of mutually commuting  {local} observables using the simplest quantum network- the bilocality scenario involving two edge parties and a central party. Each of the edge parties performs $2^{m-1}$ number of binary-outcome measurements and the central party performs $m$ number of binary-outcome measurements, where $m$ is arbitrary. We propose a suitable bilocal inequality in an arbitrary input scenario and by using  a dimension-independent approach, we derive the  optimal quantum violation of bilocality inequality. The optimal quantum value can only be obtained when the central party performs the measurements of $m$ commuting observables. This then self-tests an unbounded number of mutually commuting  {local} observables.  
 
 The plan of the paper is the following. In Sec. \ref{II}, we consider the well-known star network and show that for an even number of edge parties, the optimal quantum violation self-tests set of commuting  {local}  observables.  { 
In Sec. \ref{III}, we consider the simplest bilocal network where the central party Bob performs three measurements (i.e., $m=3$).} We demonstrate the optimal quantum violation of bilocality inequality self-tests the set of three mutually  commuting  {local}  observables. In Sec \ref{IV}, we show that this feature is generic and valid for any arbitrary $m$ input case. Considering the similar bilocality scenario, we have shown that by increasing the number of inputs for each party, one can self-test a set of an arbitrary number of mutually commuting  {local} observables. In Sec \ref{V}, we summarize the results and conclude by stating a few interesting open questions.    
\section{Self-testing of two commuting local observables in star-network }
\label{II}
\begin{figure}\begin{center}

\includegraphics[scale=0.4]{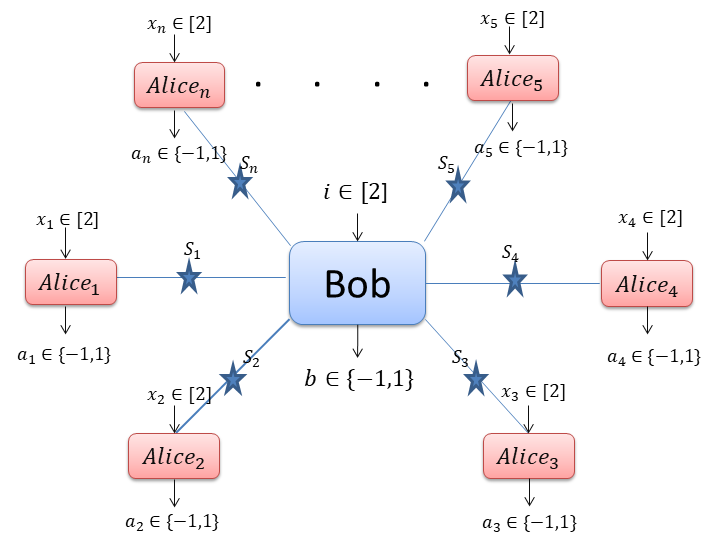}
				\caption{ Two-input-
arbitrary-party star-network}\end{center}
			\end{figure}
Let us first consider the $n$-local configuration \cite{Armi2014}   featuring  an arbitrary  $n$ number  of edge party (Alices), say Alice$_k,$  $k\in[n]$, the central party Bob and $n$ independent sources  $S_{k}$. Each  Alice$_k$  measures two binary-outcome measurements $A^{k}_{1}$ and $A^{k}_{2}$ according to the  inputs  $x_k=1$ and  $2$ respectively  and gets outputs $ {a_{k}\in \{-1,1\}}$. Bob, upon receiving the inputs $i=1,2$, performs two binary-outcome measurements $B_1$ and $B_2$  on the joint system he receives from $n$ sources and obtains output $ {b\in \{-1
,1\}}$.  The complete independence of the resources $S_{k}$s constitutes the assumption of $n$-locality which is the most crucial assumption in this context \cite{Armi2014}. 

In a $n$-local model, we assume that the hidden variables $\lambda_{k}$s, corresponding to the sources $S_{k}$ distributed according to the probability density functions $\rho_k(\lambda_{k})$s, are independent to each other. Hence, the joint distribution  $\rho{(\lambda_{1},\lambda_{2},\cdots \lambda_{n} )}$ can be written in a  factorized form as  
	\ba
	\label{fac}
	\rho{(\lambda_{1},\lambda_{2},\cdots \lambda_{n})} =\prod\limits_{k=1}^n \rho_{k}{(\lambda_{k})}
	\ea 
	which is the $n$-locality condition. Here, for  each $k\in[n]$, $\rho_{k}{(\lambda_{k})}$   satisfies the normalization condition  $\int d\lambda_{k}\rho_{k}{(\lambda_{k})}=1$. Using the $n$-locality condition for a star-network scenario, the  joint probability distribution  can be written as 
	\begin{eqnarray}\label{facn}
		&&P(a_{1}, a_{2}, \cdots , a_n, b,|x_{1},x_{2},\cdots x_n, i)\\
		\nonumber
	&=&\int\bigg(\prod\limits_{k=1}^n \rho_{k}{(\lambda_{k})}\hspace{3pt} d\lambda_{k}\hspace{3pt}  P(a_{k}|x_{k},\lambda_{k})\bigg)\times P(b|i,\lambda_{1},\lambda_{2}\cdots \lambda_{n}).
	\end{eqnarray}
	 Clearly,
	Alice$_{k}$'s outcome  solely depends on $\lambda_{k}$, but Bob's outcome depends on all of the  $\lambda_{k}$s, where $k\in[n]$. A  suitable $n$-locality inequality was proposed in \cite{Sneha2021,snehachsh}, which  is given by 
	\ba   
	\label{deltan3}(\Delta_{2}^{n})_{\color{blue}{{n-l}}}=|{I}^{n}_{2,1}|^{\frac{1}{n}}+|{I}^{n}_{2,2}|^{\frac{1}{n}}\leq 2
	\ea
 where   {'$n-l$'} denotes $n$-locality. Here $I^{n}_{2,1}$ and $I^{n}_{2,2}$ are the linear combinations of suitably chosen correlations, defined as
 
	\begin{eqnarray}
	\label{pncn31}\nonumber I^{n}_{2,1}&=&\langle(A^{1}_{1}+A^{1}_{2})(A^{2}_{1}+A^{2}_{2})\cdots (A^{n}_{1}+A^{n}_{2})B_{1}\rangle
	\\ I^{n}_{2,2}&=&\langle(A^{1}_{1}-A^{1}_{2})(A^{2}_{1}-A^{2}_{2})\cdots (A^{n}_{1}-A^{n}_{2})B_{2}\rangle
	\end{eqnarray}
	where $A^{k}_{1} (A^{k}_{2})$ denotes observables corresponding to the input $x_{k}=1(2)$ of the $k^{th}$ Alice and 
	\ba
	\langle{A^{1}_{x_1}\cdots A^{n}_{x_{n}}B_{i}}\rangle = \sum\limits_{a_{1},\cdots, a_{n},b}(-1)^{\sum_{k=1}^{n}a_{k}+b}P(a_{1},\cdots, a_{n},b|x_{1},
 \cdots, x_{n},i)
	\ea 
	We  define the expectation value of the observable of Alice$_k$ corresponding to the input $x_k$ as $\langle{A^{k}_{x_{k}}}\rangle_{\lambda_{k}} = \sum\limits_{a_{k}}(-1)^{a_{k}}  P(a_{k}|x_{k},\lambda_{k})$ where $k\in[n]$ and $x_{k}\in [2]$. Using the fact that 
	 $|\langle{B_{1}}\rangle_{\lambda_{1},\cdots \lambda_{n}}|\leq{1},$ and the sources are   independent, we can write
	\begin{eqnarray}
	\label{facn1}
|I^{n}_{2,1}|&\leq&|\langle(A^{1}_{1}+A^{1}_{2})(A^{2}_{1}+A^{2}_{2})\cdots (A^{n}_{1}+A^{n}_{2})\rangle|
	\\\label{facn2} |I^{n}_{2,2}|&\leq&|\langle(A^{1}_{1}-A^{1}_{2})(A^{2}_{1}-A^{2}_{2})\cdots (A^{n}_{1}-A^{n}_{2})\rangle|
	\end{eqnarray}
	 For simplicity let  $|A^{k}_{1}+A^{k}_{2}|=z^1_k,|A^{k}_{1}-A^{k}_{2}|=z^2_k$.    Now, using the inequality 
	\ba\label{z1}\bigg(\prod\limits_{k=1}^{n}z_{k}^{1}\bigg)^{\frac{1}{n}}+\bigg(\prod\limits_{k=1}^{n}z_{k}^{2}\bigg)^{\frac{1}{n}}\leq\prod\limits_{k=1}^{n} \bigg(z_{k}^{1}+z_k^2\bigg)^{\frac{1}{n}}\ea
	we get 
	     \begin{eqnarray}
		\label{deltan3pnc}
		(\Delta^{n}_{2})_{\color{blue}{{n-l}}}&\leq&\prod\limits_{k=1}^{n}\bigg(|A^{k}_{1}+A^{k}_{2}|+|A^{k}_{1}-A^{k}_{2}|\bigg)^{\frac{1}{n}}\hspace{1pt}
		\end{eqnarray}
Since each observable is dichotomic, clearly $|A^{k}_{1}+A^{k}_{2}|+|A^{k}_{1}-A^{k}_{2}|\leq 2, \forall k\in[n]$. Hence 
	    we finally obtain $(\Delta^{n}_{2})_{\color{blue}{{n-l}}}\leq 2$ as claimed in Eq. (\ref{deltan3}).
		
 	 {To derive the optimal quantum value of $(\Delta^{n}_{2})_{Q}$, we use the following approach. Without loss of serious generality, we consider the following state $|\psi\rangle =\otimes_{k=1}^n|\psi\rangle_{A_kB}$ and define two suitable vectors $M^{n}_{2,1}|\psi\rangle $ and $M^{n}_{2,2}|\psi\rangle$  as follows:}  
  {\ba\na M^{n}_{2,1}|\psi\rangle&=&\left[\bigotimes\limits_{k=1}^{n}\left(\frac{{A}^{k}_{1}+{A}^{k}_{2}}{(\omega_{2,1}^{n})_{A_{k}}}\right)\otimes B_1\right]|\psi\rangle,\\
M^{n}_{2,2}|\psi\rangle&=&\left[\bigotimes\limits_{k=1}^{n}\left(\frac{{A}^{k}_{1}-{A}^{k}_{2}}{(\omega_{2,2}^{n})_{A_{k}}}\right)\otimes B_2\right]|\psi\rangle\label{Mn2}
 \ea}
 {Here $|\psi\rangle_{A_kB}$ is the state shared between Alice$_k$ and Bob. Here  $(\omega^{n}_{2,i})_{A_{k}}$  is the norm of   the  vector $\left[{A}^{k}_{1}+(-1)^i{A}^{k}_{2}\right]|\psi\rangle_{A_kB}$ such that each of   the  vector $\frac{\left[{A}^{k}_{1}+(-1)^i{A}^{k}_{2}\right]|\psi\rangle_{A_kB}}{(\omega^{n}_{2,i})_{A_{k}}}$ becomes normalized which in turn provides that     $M^{n}_{2,i}|\psi\rangle$ is  normalized.  Hence we get } 
 
 {\ba\label{I2qm}
I^{n}_{2,1}&=& \omega_{2,1}^{n}\langle M^{n}_{2,1}\rangle, \quad 
I^{n}_{2,2}=\omega_{2,2}^{n} \langle M^{n}_{2,2}\rangle\ea}
 {where $\omega_{2,1}^{n}=\prod\limits_{k=1}^n(\omega_{2,1}^{n})_{A_k}$ and $\omega_{2,2}^{n}=\prod\limits_{k=1}^n(\omega_{2,2}^{n})_{A_k}$.  {Since $(\omega_{2,1}^{n})$ and $(\omega_{2,2}^{n})$  are products of norms, these are always positive}. Hence from Eq. (\ref{I2qm}), we can write}
 {\ba
\label{12}
(\Delta^{n}_{2})_{Q} &=&\left(\omega_{2,1}^{n}\big|\langle M^{n}_{2,1}\rangle\big|\right)^{1/n}+\left(\omega_{2,2}^{n}\big|\langle M^{n}_{2,2}\rangle\big|\right)^{1/n}\ea }
 { From Eq. (\ref{12}), it is straightforward to argue that the optimal value of $(\Delta^{n}_{2})_{Q}$ is obtained when $\langle M^{n}_{2,1}\rangle=\pm 1$ and $\langle M^{n}_{2,2}\rangle=\pm 1$ hold. This ensures that the quantum state shared by each Alice and Bob has to be a pure state $|\psi\rangle$  which is the eigenvector of both of $M^{n}_{2,1}$ and $M^{n}_{2,2}$ corresponding to the eigenvalue $\pm 1$. 
 i.e.,  
$M^{n}_{2,1}|\psi\rangle=\pm |\psi\rangle$ and $M^{n}_{2,2}|\psi\rangle=\pm|\psi\rangle$.} 
This implies that
	\begin{eqnarray}(\Delta^{n}_{2})_{Q}^{opt} =\max_{\{A^k_1,A^k_2\}}\big[(\omega^{n}_{2,1})^{\frac{1}{n}}+(\omega^{n}_{2,2})^{\frac{1}{n}}\big]
	\end{eqnarray}
 {where $(\omega_{2,1}^n)_{A_k}$ and $(\omega_{2,2}^n)_{A_k}$ are given by} 
 { \begin{eqnarray}
\label{omega12}
 \nonumber
(\omega^{n}_{2,1})_{A_{k}}=||(A^{k}_{1}+A^{k}_{2})|\psi\rangle_{A_kB}||_{2}=\sqrt{2+\langle \{A^{k}_{1},(A^{k}_{2}\}\rangle}\\
(\omega^{n}_{2,2})_{A_{k}}=||(A^{k}_{1}-A^{k}_{2})|\psi\rangle_{A_kB}||_{2}=\sqrt{2-\langle \{A^{k}_{1},(A^{k}_{2}\}\rangle}
 \end{eqnarray}}
 By using the  inequality  Eq. (\ref{z1}),
	we get 
  \ba\label{wn3i}(\omega^{n}_{2,1})^{\frac{1}{n}}+(\omega^{n}_{2,2})^{\frac{1}{n}}&\leq&\prod\limits_{k=1}^{n}\bigg( (\omega^{n}_{2,1})_{A_{k}}+(\omega^{n}_{2,2})_{A_{k}}\bigg)^{\frac{1}{n}}\\\na
&=& \prod\limits_{k=1}^{n}\bigg(
\sqrt{2+\langle \{A^{k}_{1},A^{k}_{2}\}\rangle}+\sqrt{2-\langle \{A^{k}_{1},A^{k}_{2}\}\rangle}\bigg)^{\frac{1}{n}}\\\na
&\leq& \prod\limits_{k=1}^{n}\bigg(\sqrt{4+2\sqrt{4-\langle \{A^{k}_{1},A^{k}_{2}\}\rangle^2}}\bigg)^{\frac{1}{n}}
\ea
 
	Hence, to obtain the optimal value of  $(\Delta^{n}_{2})_{Q}^{opt}$, observables of each Alice$_k$ has to  be mutually  anticommuting, i.e., $\{A^{k}_{1},A^{k}_{2}\}=0$. This provides the optimal quantum value is   $(\Delta^n_2)^{opt}_Q=2\sqrt{2}$. 
 {Considering the optimal scenario, from Eq. (\ref{omega12}), we then get $(\omega^{n}_{2,1})_{A_k}=(\omega^{n}_{2,2})_{A_k}=\sqrt{2}$.  Let us denote $\mathcal{A}^n_1=\bigotimes\limits_{k=1}^{n}\left(\frac{{A}^{k}_{1}+{A}^{k}_{2}}{\sqrt{2}}\right)$ and $\mathcal{A}^n_2=\bigotimes\limits_{k=1}^{n}\left(\frac{{A}^{k}_{1}-{A}^{k}_{2}}{\sqrt{2}}\right)$. It is then easy to check that when the optimal quantum value $(\Delta^n_2)^{opt}_Q=2\sqrt{2}$ is achieved, we have \ba\label{2A}\mathcal{A}^n_1\mathcal{A}^n_2=(-1)^n\mathcal{A}^n_2\mathcal{A}^n_1.\ea}
 {We can then write $M^{n}_{2,1}=\mathcal{A}_{1}^{n}\otimes B_1$ and $ M^{n}_{2,2}=\mathcal{A}_{2 }^{n}\otimes B_2$.
 Since  $M^{n}_{2,1}|\psi\rangle=\pm|\psi\rangle$ and $M^{n}_{2,2}|\psi\rangle=\pm |\psi\rangle$, the observables $M^{n}_{2,1}$  and $M^{n}_{2,2}$ are commuting, i.e., $\left[M^{n}_{2,1},M^{n}_{2,2}\right]=0$. 
Using Eq. (\ref{2A}), we get}
 {\ba\mathcal{A}^{n}_1\mathcal{A}^{n}_2\otimes \left(B_1B_2-(-1)^n B_2B_1\right)=0
\ea}
  This implies that for even $n$, the observables $B_1$ and $B_2$ need to be commuting. In other words, the optimal quantum value $(\Delta^{n}_{2})^{opt}_{Q}$  self-tests the commuting observables - an interesting self-testing that has not been explored earlier.  
  It is important to note that the above derivation is dimension-independent and hence the conclusion holds for any  dimensional system. 

However, a realization of such observables for each Alice$_{k}$ can be found even for the local qubit system as follows:
\ba\label{obs} A^k_1=\frac{\sigma_z+\sigma_x}{\sqrt{2}}, \quad A^k_2=\frac{\sigma_z-\sigma_x}{\sqrt{2}}, \forall k\in[n].\ea
 Hence  Bob's the observables $B_1$ and $B_2$ are 
 $$B_1=\otimes^n \sigma_z, \quad B_2=\otimes^n \sigma_x$$
 which in turn provides that  
$B_1B_2=(-1)^n B_2B_1$, i.e., for an even value of $n$, the observales $B_1$ and $B_2$ are commuting. It is straightforward to show that each edge party shares a two-qubit entangled state with the central party, Bob. Hence, the optimal quantum violation of the inequality (\ref{deltan3}) self-tests two mutually commuting observables iff the star-network is featuring an even number of parties.
 \section{Self-testing a set of three mutually commuting {local} observables}
 \label{III} 
We now focus on self-testing three mutually commuting  {local} observables by considering the simplest bilocal scenario. Further, we generalize this result for an arbitrary input bilocal scenario for self-testing an unbounded number of mutually commuting  {local} observables. 

As depicted in Fig.2, we consider the bilocal scenario featuring two edge parties Alice and Charlie, and a central party Bob. Alice and Charlie,  each share a physical state with Bob generated from two independent sources, $S_1$ and  $S_2$ respectively. The central party, Bob, performs three dichotomic measurements $B_1, B_2$ and $B_3$ according to the inputs $i\in[3]$ and obtains the outcome $ {b\in\{-1,1\}}$. Alice (Charlie) receives inputs  $x\in[4] (z\in[4])$  
 and performs four  measurements $A_{x} (C_z)$ accordingly   and obtains outputs $ {a(c)\in \{-1,1\}}$.

   If the joint probability distribution $P(a,b,c|x,i,z)$
can be factorized as \ba\label{factm3}\na
		P(a,b,c|x,i,z)=&\int\rho_1(\lambda_{1}) \rho_2(\lambda_{2})& P(a|x,\lambda_{1}) \hspace{1mm}P(b|i,\lambda_{1},\lambda_{2}) \hspace{1mm}\\&& \hspace{1mm}P(c|z,\lambda_{2})\hspace{1mm}
 d\lambda_{1} d\lambda_{2}\ea
where $\lambda_1$ and $\lambda_2$ are the physical states generated from the source $S_1$ and $S_2$ respectively, then   we propose that the following inequality 
	\ba  \label{Adeltan3}
	(\Delta_{3}^{2})_{ {b-l}}=\sqrt{|{I}^{2}_{3,1}|}+\sqrt{|{I}^{2}_{3,2}|}+\sqrt{|{I}^{2}_{3,3}|}\leq 6
	\ea
is satisfied. Here   {`$b-l$} ' denotes bi-locality. Here, $I^{2}_{3,1}$  $I^{2}_{3,2}$, and $I^{2}_{3,3}$  are the linear combinations of suitably chosen correlations defined as 
	\begin{eqnarray}
	\label{Apncn31}\nonumber
	I^{2}_{3,1}&=&\bigg\langle(A_{1}+A_{2}+A_{3}-A_{4}) B_{1}(C_{1}+C_{2}+C_{3}-C_{4})\bigg\rangle
	\\ \hspace{2mm}
I^{2}_{3,2}&=&\bigg\langle(A_{1}+A_{2}-A_{3}+A_{4}) B_{2}(C_{1}+C_{2}-C_{3}+C_{4})\bigg\rangle\\\nonumber
	I^{2}_{3,3}&=&\bigg\langle(A_{1}-A_{2}+A_{3}+A_{4}) B_{3}(C_{1}-C_{2}+C_{3}+C_{4})\bigg\rangle
	\end{eqnarray}
	 where the expectation value is defined as
	\ba
	\nonumber
	\langle{A^{1}_{x}B_{i}C_z}\rangle = \sum\limits_{a,b,c}(-1)^{a+b+c}P(a,b,c|x,i,z)
	\ea 
	By defining $\langle{A_{x}}\rangle_{\lambda_1} = \sum\limits_{a}(-1)^{a}  P(a|x,\lambda_1)$ where $x\in [4]$ and using the facts that  $|\langle{B_{1}}\rangle_{\lambda_{1},\lambda_{2}}|\leq{1},$ and the sources are   independent, we can write
	\begin{eqnarray}
	\label{Afacn31}\nonumber
	|I^{2}_{3,1}|&\leq&\big|\big\langle(A_{1}+A_{2}+A_{3}-A_{4})\big\rangle\big|\big|\big\langle(C_{1}+C_{2}+C_{3}-C_{4})\big\rangle\big|\\
	\end{eqnarray}
	Similarly, we can  factorize $|I^{2}_{3,2}|$ and $|I^{2}_{3,3}|$ as Eq. (\ref{Afacn31}).   Now, using the inequality \begin{equation}
	\label{ATavakoli}
	\ \forall\ \ z_{k}^{i} \geq 0; \ \ \ \sum\limits_{i=1}^{m}\bigg(\prod\limits_{k=1}^{n}z_{k}^{i}\bigg)^{\frac{1}{n}}\leq \prod \limits_{k=1}^{n}\bigg(\sum\limits_{i=1}^{m}z_{k}^{i}\bigg)^{\frac{1}{n}}
	\end{equation} for $m=3$ and $n=2$, we get 
	\begin{eqnarray}\label{Am=3}\na
	(\Delta_{3}^{2})_{ {b-l}}&\leq&\bigg(\big|\big\langle(A_{1}+A_{2}+A_{3}-A_{4})\big\rangle\big|\big|\big\langle(C_{1}+C_{2}+C_{3}-C_{4})\big\rangle\big|\bigg)^{1/2}\\\na&&+\bigg(\big|\big\langle(A_{1}+A_{2}-A_{3}+A_{4})\big\rangle\big|
 \big|\big\langle(C_{1}+C_{2}-C_{3}+C_{4})\big\rangle\big|\bigg)^{1/2}\\\na&&+\bigg(\big|\big\langle(A_{1}-A_{2}+A_{3}+A_{4})\big\rangle\big|\big|\big\langle(C_{1}-C_{2}+C_{3}+C_{4})\big\rangle\big|\bigg)^{1/2}\\\na
 &\leq & \bigg(\big|\big\langle(A_{1}+A_{2}+A_{3}-A_{4})\big\rangle\big|+\big|\big\langle(A_{1}+A_{2}-A_{3}+A_{4})\big\rangle\big|\\\na&&+\big|\big\langle(A_{1}-A_{2}+A_{3}+A_{4})\big\rangle\big|\bigg)^{1/2}\times\bigg(\big|\big\langle(C_{1}+C_{2}+C_{3}-C_{4})\big\rangle\big|\\\na&&+\big|\big\langle(C_{1}+C_{2}-C_{3}+C_{4})\big\rangle\big|+\big|\big\langle(C_{1}-C_{2}+C_{3}+C_{4})\big\rangle\big|\bigg)^{1/2}\\
\end{eqnarray}
Let us denote   
$\big|\big\langle(A_{1}+A_{2}+A_{3}-A_{4})\big\rangle\big|+\big|\big\langle(A_{1}+A_{2}-A_{3}+A_{4})\big\rangle\big|+\big|\big\langle(A_{1}-A_{2}+A_{3}+A_{4})\big\rangle\big|=\eta^A_3,
\big|\big\langle(C_{1}+C_{2}+C_{3}-C_{4})\big\rangle\big|+\big|\big\langle(C_{1}+C_{2}-C_{3}+C_{4})\big\rangle\big|+\big|\big\langle(C_{1}-C_{2}+C_{3}+C_{4})\big\rangle\big|=\eta^C_3
$. 

Since each observable is dichotomic,  we get $\eta^A_3=\eta^C_3\leq 6.$   Substituting these  in Eq.(\ref{Am=3}), we finally obtain  $(\Delta^{2}_{3})_{ {b-l}}\leq 6$ as claimed in Eq. (\ref{Adeltan3}).

	
	 {To derive the optimal quantum value of $(\Delta^{2}_{3})_{Q}$, we use the following approach. Without loss of serious generality, we consider the following state $|\psi\rangle =|\psi\rangle_{AB}\otimes |\psi\rangle_{BC}$ and the suitable vectors   $M^{2}_{3,1}|\psi\rangle, M^{2}_{3,2}|\psi\rangle$ and $M^{2}_{3,3}|\psi\rangle$  as follows:}  
 
	 {\ba
	\label{M23}\na
M^{2}_{3,1}|\psi\rangle&=&\bigg(\frac{A_1+A_2+A_3-A_4}{(\omega_{3,1}^{2})_A}\otimes\frac{C_1+C_2+C_3-C_4}{(\omega_{3,1}^{2})_C}\otimes B_1\bigg)|\psi\rangle\\\na M^{2}_{3,2}|\psi\rangle&=&\bigg(\frac{A_1+A_2-A_3+A_4}{(\omega_{3,2}^{2})_A}\otimes\frac{C_1+C_2-C_3+C_4}{(\omega_{3,2}^{2})_C}\otimes B_2\bigg)|\psi\rangle\\\na
M^{2}_{3,3}|\psi\rangle&=&\bigg(\frac{A_1-A_2+A_3+A_4}{(\omega_{3,3}^{2})_A}\otimes\frac{C_1-C_2+C_3+C_4}{(\omega_{3,3}^{2})_C}\otimes B_3\bigg)|\psi\rangle\\
	\ea}
  { Here $|\psi\rangle_{AB}(|\psi\rangle_{BC})$ is the state shared between Alice (Charlie) and Bob. Also  $(\omega^{2}_{3,1})_{A}$  is the norm of the  vector $({A}_{1}+A_2+A_3-A_4)|\psi\rangle_{AB}$ such that  the vector $\frac{({A}_{1}+A_2+A_3-A_4)|\psi\rangle_{AB}}{(\omega^{2}_{3,1})_{A}}$ becomes normalized. The similar argument holds for each $(\omega^{2}_{3,i})_{A/C}, i\in[3]$.   This in turn provides that  the vectors $M^2_{3,i}|\psi\rangle$ s are also normalized. }

 {Using the observables of Eq. (\ref{M23}), we can write }	
 {\ba\label{I3qm}\na
I^{2}_{3,1}&=& \omega_{3,1}^{2}\langle M^{2}_{3,1}\rangle;\quad
I^{2}_{3,2}=\omega_{3,2}^{2} \langle M^{2}_{3,2}\rangle; \quad 
I^{2}_{3,3}=\omega_{3,3}^{2}\langle M^{2}_{3,3}\rangle\\
\ea}
 {where $\omega^{2}_{3,i}$ is defined as    $\omega^{2}_{3,i}=(\omega^{2}_{3,i})_{A}(\omega^{2}_{3,i})_{C}$.  Since $(\omega_{3,i}^{2})$s  are products of norms, these are always positive. Hence from Eq. (\ref{I3qm}), we can write}
 {\ba\na
(\Delta^{2}_{3})_{Q}&=&\sqrt{\omega_{3,1}^{2}|\langle M^{2}_{3,1}\rangle|}+\sqrt{\omega_{3,2}^{2}|\langle M^{2}_{3,2}\rangle|}+\sqrt{\omega_{3,3}^{2}|\langle M^{2}_{3,3}\rangle|}
\ea}

 { As we have defined, the vectors $M^{2}_{3,1}|\psi\rangle, M^{2}_{3,2}|\psi\rangle$ and $M^{2}_{3,3}|\psi\rangle$ are normalized,  the optimal value of $(\Delta^{2}_{3})_{Q}$ is obtained when  the conditions  $\langle M^{2}_{3,1}\rangle=\pm 1,\langle M^{2}_{3,2}\rangle=\pm 1 $ and $\langle M^{2}_{3,3}\rangle=\pm 1$ holds simultaneously. This ensures that the state quantum state    $|\psi\rangle$ has to be a pure state, and it  is the eigenvector of each of the   observable  $M^{2}_{3,1},M^{2}_{3,2} $ and $M^{2}_{3,3}$ corresponding to the eigenvalue $\pm1$ 
 i.e.,  
$M^{2}_{3,1}|\psi\rangle=\pm |\psi\rangle, M^{2}_{3,2}|\psi\rangle=\pm |\psi\rangle$ and $M^{2}_{3,3}|\psi\rangle=\pm |\psi\rangle$.} 
This implies that 
	\begin{eqnarray}(\Delta^{2}_{3})^{opt}_{Q}=\max_{A_x,C_z}\left(\sqrt{\omega^{2}_{3,1}}+\sqrt{\omega^{2}_{3,2}}+\sqrt{\omega^{2}_{3,3}}\right)
	\end{eqnarray}
 {The norm $(\omega^{2}_{3,1})_A$ is given by  }
 {\ba
	&&(\omega^{2}_{3,1})_{A}=||(A_{1}+A_{2}+A_{3}-A_{4})|\psi\rangle_{AB}||_{2}\\\na
	&=&\Big(4+\langle \{A_{1},(A_{2}+A_{3}-A_{4})\}+\{A_{2},(A_{3}-A_{4})\}-\{A_{3},A_{4}\}\rangle\Big)^{1/2}	\ea}
	 {Similarly,  we can write for
	$(\omega^{2}_{3,2})_{A}((\omega^{2}_{3,2})_{C})$ and $(\omega^{2}_{3,3})_{A}((\omega^{2}_{3,3})_{C})$.}
	 Since  $\omega^{2}_{3,i}=(\omega^{2}_{3,i})_{A}(\omega^{2}_{3,i})_{C}, \forall i\in[3]$, by using the  inequality  Eq. (\ref{ATavakoli}),
	we get that
 \ba \label{Awn3i}\sum\limits_{i=1}^3\sqrt{\omega^{2}_{3,i}}&\leq&\sqrt{\sum\limits_{i=1}^3(\omega^{2}_{3,i})_A}\sqrt{\sum\limits_{i=1}^3(\omega^{2}_{3,i})_C}\ea
	Further using the convex inequality, we have  \ba\label{Aconvex3}\sum\limits_{i=1}^3(\omega^{2}_{3,i})_A\leq \sqrt{3\sum\limits_{i=1}^3\bigg[(\omega^{2}_{3,i})_A\bigg]^2}\ea 

The equality holds when each of $(\omega^{n}_{3,i})_{A}$ is equal. We can then write  	\begin{eqnarray}\label{Asumw23ik}\na\sum\limits_{i=1}^3\bigg[(\omega^{2}_{3,i})_A\bigg]^2&=& \langle\psi|12+(\{A_{1},(A_{2}+A_{3}+A_{4})\}-\{A_{2},(A_{3}+A_{4})\}\\&&-\{A_{3},A_{4}\})|\psi\rangle=\langle\psi(12+\delta_3)|\psi\rangle
		\end{eqnarray} where 
  \ba\label{del3main}\delta_3=(\{A_{1},(A_{2}+A_{3}+A_{4})\}-\{A_{2},(A_{3}+A_{4})\}-\{A_{3},A_{4}\})\ea

		Let $|\psi'\rangle=(A_{1}-A_{2}-A_{3}-A_{4})|\psi\rangle$  such that $|\psi\rangle\neq 0$. Then
		$\langle\psi'|\psi'\rangle=\langle\psi|(4-\delta_3)|\psi\rangle$ implies $\langle\delta_3\rangle=4-\langle\psi'|\psi'\rangle$.
Evidently, $\langle\delta_3\rangle_{max}$ is obtained only when $\langle\psi'|\psi'\rangle=0$. Since, $|\psi\rangle\neq 0$, we then have 
\ba\label{Ac3}A_{1}-A_{2}-A_{3}-A_{4}=0\ea
	Hence, to obtain the optimal value of  $(\Delta^{2}_{3})_{Q}^{opt}$,   the observables of Alice must satisfy  the linear condition of Eq.(\ref{Ac3}). In turn  $\langle\delta_3\rangle_{max}=4$  provides $((\omega^{2}_{3,1})_A)^2+((\omega^{2}_{3,2})_A)^2+((\omega^{2}_{3,3})_A)^2=16$. Plugging it in the above mentioned  inequality (\ref{Aconvex3}), we get $\sum\limits_{i=1}^3(\omega^{2}_{3,i})_A\leq 4\sqrt{3}$.  Since each observable $A_{x}$ is dichotomic, pre-multiplying and post-multiplying Eq.(\ref{Ac3}) by $A_1$, and adding them we get
	 
	 \ba\label{AAK3}\{A_1,A_2\}+\{A_1,A_3\}+\{A_1,A_4\}=2\ea
	 Similarly, we can find three more such relations, which  are the following:
	 \ba\label{Ac3d}\nonumber
	 \{A_1,A_2\}-\{A_2,A_3\}-\{A_2,A_4\}&=&2
	\\
	\{A_1,A_3\}-\{A_2,A_3\}-\{A_3,A_4\}&=&2\\\nonumber
	\{A_1,A_4\}-\{A_2,A_4\}-\{A_3,A_4\}&=&2\ea
	Solving  Eqs. (\ref{AAK3}-\ref{Ac3d}), we get 
 \ba\label{Aaanti3}
 &&\{A_{1}, A_{2}\}=\{A_{1}, A_{3}\}=\{A_{1}, A_{4}\}=\frac{2}{3}\\
 \nonumber
 &&\{A_{2}, A_{3}\}=\{A_{2}, A_{4}\}=\{A_{3}, A_{4}\}=-\frac{2}{3}.\ea   We thus obtain the relations between the  observables for each Alice. Also, for optimal value, we check that $(\omega^{2}_{3,1})_{A}=(\omega^{2}_{3,2})_{A}=(\omega^{2}_{3,3})_{A}=4/\sqrt{3}$. Following a similar calculation for the other party Charlie, we get that
 \ba\label{Acanti3}
 &&\{C_{1}, C_{2}\}=\{C_{1}, C_{3}\}=\{C_{1}, C_{4}\}=\frac{2}{3}\\
 \nonumber
 &&\{C_{2}, C_{3}\}=\{C_{2}, C_{4}\}=\{C_{3}, C_{4}\}=-\frac{2}{3}.\ea
 And for the optimal violation, we get  $(\omega^{2}_{3,1})_{C}=(\omega^{2}_{3,2})_{C}=(\omega^{2}_{3,3})_{C}=4/\sqrt{3}$.  This, in turn, provides the optimal quantum value 
	 \ba(\Delta^{2}_{3})_{Q}^{opt}= 4\sqrt{3}.\ea
 {Considering the optimal scenario,  let us denote} 

 {\ba\na\mathcal{A}^{  2 ,3}_1&=&\left(\frac{{A}_{1}+{A}_{2}+{A} _{3}-{A} _{4}}{4/\sqrt{3}}\right)\otimes\left(\frac{{C}_{1}+{C} _{2}+{C}_{3}-{C}_{4}}{4/\sqrt{3}}\right)\\
 \mathcal{A}^{  2,3 }_2&=&\left(\frac{{A} _{1}+{A} _{2}-{A} _{3}+{A}_{4}}{4/\sqrt{3}}\right)\otimes\left(\frac{{C}_{1}+{C}_{2}-{C} _{3}+{C}_{4}}{4/\sqrt{3}}\right)\\\na 
 \mathcal{A}^{  2,3 }_3&=&\left(\frac{{A} _{1}-{A}_{2}+{A} _{3}+{A}_{4}}{4/\sqrt{3}}\right)\otimes\left(\frac{{C}_{1}-{C}_{2}+{C}_{3}+{C}_{4}}{4/\sqrt{3}}\right)\ea}
 {Since the optimal quantum violation certifies   the relations Eq. (\ref{Aaanti3}) and Eq. (\ref{Acanti3}), we obtain that}  {\ba
\label{3A}\mathcal{A}^{2,3}_i\mathcal{A}^{2,3 }_j=\mathcal{A}^{2,3}_j\mathcal{A}^{2,3}_i, \quad \forall i\neq j\in[3]\ea} 
 
  {We can then write $M^{2}_{3,1}=\mathcal{A}_{1}^{2,3}\otimes B_1, M^{2}_{3,2}=\mathcal{A}_{2}^{2,3}\otimes B_2$ and $ M^{2}_{3,3}=\mathcal{A}_{3 }^{2,3}\otimes B_3$.
 Since  $M^{2}_{3,i}|\psi\rangle=\pm|\psi\rangle, \forall i\in[3]$, the observables $M^{2}_{3,1}$, $M^{2}_{3,2}$  and $M^{2}_{3,3}$ are mutually  commuting, i.e., $\left[M^{2}_{3,1},M^{2}_{3,2}\right]=0$. 
Using Eq. (\ref{3A}), we get}
 {\ba\mathcal{A}^{2,3}_1\mathcal{A}^{2,3}_2\otimes \left(B_1B_2-B_2B_1\right)=0
\ea}
which implies that  $B_1$ and $B_2$ are commuting. 
Following a similar procedure, it is straightforward to show that $[B_2,B_3]=[B_1, B_3]=0$.  Thus, the optimal quantum violation of the bilocality inequality (\ref{Adeltan3})  uniquely fixes Bob's observables  and  simultaneously a set of three mutually commuting  {local} observables has been  self-tested. 

In the next section, we show that this feature is generic and valid for any arbitrary $m$ input case. In other words, we demonstrate the self-testing of an unbounded number of mutually commuting  {local} observables.
    
  \section{Self-testing a set of arbitrary $m$ number of mutually commuting {local} observables}\label{IV}
	\begin{figure}[h]\begin{center}
				\includegraphics[scale=0.4]{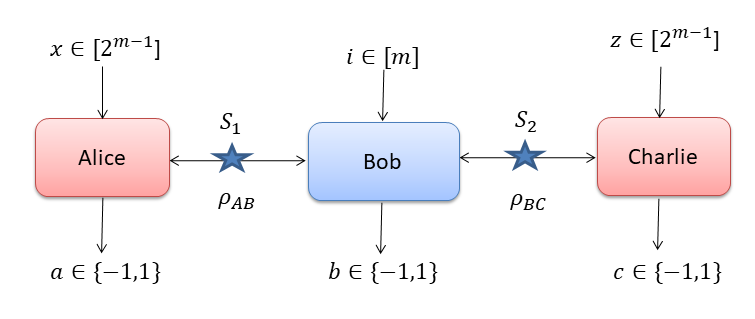}
				\caption{Arbitrary input scenario  in bilocal  network}\end{center}
			\end{figure} 
In order to self-test a set of an unbounded  $m$ number of mutually commuting  {local} observables,  we consider the bilocality scenario with arbitrary $m$ input and derive the bilocality inequality. Here the central party Bob receives arbitrary $m$ inputs and each of the edge parties  Alice and Charlie receives $2^{m-1}$ inputs. In this tripartite scenario,   the two edge parties, Alice and Charlie, receive respective inputs  $x, z\in\{1,2,3,\cdots,2^{m-1}\}$ producing outputs $ {a, c\in\{-1,1\}}$ respectively. Bob's  inputs are denoted as $i\in\{1,2,\cdots,m\}$ and produces output $ {b\in\{-1,1\}}$.
It is assumed that there are two independent sources, $S_1$ and $S_2$, and each of them distributes a state with the central party Bob. 

We propose  the generalized bilocality inequality  for arbitrary $m$ as
	\begin{equation}
	\label{deltam}
	(\Delta^{2}_{m})_{ {b-l}}=\sum\limits_{i=1}^m \sqrt{|J_{m, i}|}\leq m\binom{m-1}{\lfloor\frac{m-1}{2}\rfloor}
	\end{equation} 
	
	where $	J_{m, i}$ is the linear combinations of suitable correlations,  defined as 
		\begin{equation}
		\label{nbell}
		{J}_{m,i}=\bigg\langle\bigg(\sum\limits_{x=1}^{2^{m-1}}(-1)^{y^{x}_{i}} A_{x}\bigg)B_{i}\bigg(\sum\limits_{z=1}^{2^{m-1}}(-1)^{y^{z}_{i}} C_{z}\bigg)\bigg\rangle
	\end{equation}
	Here ${y^{x(z)}_{i}}$ takes value either $0$ or $1$.  For our purpose, we fix the values of ${y^{x}_{i}}$ and $y^{z}_{i}$ by using  the encoding scheme used in Random Access Codes (RACs) \cite{pan2020,Ambainis,Ghorai2018} as a tool. This will fix $1$ or $-1$ values of $(-1)^{y^{x}_{i}}$ in Eq. (\ref{nbell}) for a given $i$. Let us consider a random variable $y^{\alpha}\in \{0,1\}^{m}$ with $\alpha\in \{1,2...2^{m}\}$. Each element of the bit string can be written as $y^{\alpha}=y^{\alpha}_{i=1} y^{\alpha}_{i=2} y^{\alpha}_{i=3} .... y^{\alpha}_{i=m}$. For  example, if $y^{\alpha} = 011...00$ then $y^{{\alpha}}_{i=1} =0$, $y^{{\alpha}}_{i=2} =1$, $y^{{\alpha}}_{i=3} =1$ and so on. Here we  denote  the length $m$ binary strings as $y^{x}$.   Now  we consider the bit strings such that for any two $x$ and $x'$,  $y^{x}\oplus_2y^{x'}=11\cdots1$. Clearly, we have $x\in \{1,2...2^{m-1}\}$ constituting the inputs for Alice. If $i=1$, we get all the first bit of each  bit  string $y^x$ for every $x\in \{1,2 \cdots 2^m\}$.
This similar encoding holds for the other party, Charlie. 		
	
	An example for $m=2$ could be useful here. In this case, we have $y^{\delta}\in \{00,01,10,11\}$ with $\delta =1,2,3,4$. We then denote $y^{x} \equiv \{y^{1},y^{2}\} \in \{00,01\}$ with $y^{1}=00$ and $y^{2}=01$. This also means  $y^{1}_{i=1}=0, y^{1}_{i=2}=0, y^{2}_{i=1}=0$ and $y^{2}_{i=2}=1$.

	Here, using the fact that the observable $|B_i|_{\lambda_1,\lambda_2}\leq 1$,  we get 
\begin{equation}
		\label{m2}
		|{J}_{m,i}|\leq \bigg|\bigg\langle\sum\limits_{x=1}^{2^{m-1}}(-1)^{y^{x}_{i}} A_{x}\bigg\rangle\bigg|\bigg|\bigg\langle\sum\limits_{z=1}^{2^{m-1}}(-1)^{y^{z}_{i}} C_{z}\bigg)\bigg\rangle\bigg|
	\end{equation}
This implies that 
\ba \sum\limits_{i=1}^m\sqrt{|{J}_{m,i}|}\leq\sum\limits_{i=1}^m\sqrt{\bigg|\bigg\langle\sum\limits_{x=1}^{2^{m-1}}(-1)^{y^{x}_{i}} A_{x}\bigg\rangle\bigg|\bigg|\bigg\langle\sum\limits_{z=1}^{2^{m-1}}(-1)^{y^{z}_{i}} C_{z}\bigg)\bigg\rangle\bigg|}\ea
Using the inequality
\ba\label{z2}\sum\limits_{i=1}^{m}\sqrt{z^i_{1}z^i_2}\leq\sqrt{\bigg(\sum\limits_{i=1}^{m}z^i_1\bigg)\bigg(\sum\limits_{i=1}^{m}z^i_2\bigg)}\ea
we get 
		\begin{eqnarray}\na\label{deltanmleq}
	(\Delta_{m}^{2})_{ {b-l}} &\leq&\sqrt{\bigg(\sum\limits_{i=1}^{m} \bigg|\bigg\langle\sum\limits_{x=1}^{2^{m-1}}(-1)^{y^{x}_{i}} A_{x}^{}\bigg\rangle\bigg|\bigg)\bigg(\sum\limits_{i=1}^{m} \bigg|\bigg\langle\sum\limits_{z=1}^{2^{m-1}}(-1)^{y^{z}_{i}} C_{z}^{}\bigg\rangle\bigg|\bigg)}\\
 &\leq &\sqrt{\eta^A_m\times \eta^C_m}
		\end{eqnarray}
where 
     \ba\label{etaAC}\eta^A_m=\sum\limits_{i=1}^{m}\bigg|\bigg\langle\sum\limits_{x=1}^{2^{m-1}}(-1)^{y^{x}_{i}} A_{x}\bigg\rangle\bigg|,\quad \eta^C_m=\sum\limits_{i=1}^{m}\bigg|\bigg\langle\sum\limits_{z=1}^{2^{m-1}}(-1)^{y^{z}_{i}} C_{z}\bigg\rangle\bigg| \ea
Since the encoding scheme used for Alice and Charlie is identical, clearly  \ba\label{etam}(\eta^A_m)_{max}=(\eta^C_m )_{max}\ea
which is the optimal bilocal bound of $\Delta_m^2$.
We have obtained that \ba\label{etamA}(\eta^A_m)_{max}=m\binom{m-1}{\lfloor\frac{m-1}{2}\rfloor}\ea The detailed calculation is provided in Appendix \ref{A}. Substituting this in Eq.(\ref{deltanmleq}), we get back the bilocality inequality as defined in 
    Eq. (\ref{deltam}).


 {To obtain  the quantum upper bound of the expression $(\Delta_{m}^{2})$,  we use the following approach. To derive the optimal quantum value of $(\Delta^{2}_{3})_{Q}$, we use the following approach. Without loss of serious generality, we consider the following state $|\psi\rangle =|\psi\rangle_{AB}\otimes |\psi\rangle_{BC}$ and the suitable vectors $M^{2}_{m,i}|\psi\rangle$ as follows:}  
  {\ba\na
	\label{M2m}
M^{2}_{m,i}|\psi\rangle&=\Bigg(&\bigg(\frac{1}{(\omega_{m,i}^{2})_A}\sum\limits_{x=1}^{2^{m-1}}(-1)^{y^{x}_{i}} A_{x}\bigg)\otimes B_i\\&&\hspace{3mm}\otimes\hspace{1mm}  \bigg(\frac{1}{(\omega_{m,i}^{2})_C}\sum\limits_{z=1}^{2^{m-1}}(-1)^{y^{z}_{i}} 
C_{z}\bigg)\hspace{3mm}\Bigg)|\psi\rangle\ea}
 {where $(\omega^{2}_{m,i})_{A}$  is the norm of the  vector $(\sum\limits_{x=1}^{2^{m}-1}(-1)^{y^{x}_{i}} A_{x})|\psi\rangle_{AB}$ such that  the vector $(\sum\limits_{x=1}^{2^{m}-1}(-1)^{y^{x}_{i}} A_{x})|\psi\rangle_{AB}/(\omega^{2}_{m,i})_{A}$ becomes normalized. The similar argument holds for each $(\omega^{2}_{m,i})_{C}, i\in[m]$.   This in turn provides that  the vectors $M^2_{m,i}|\psi\rangle_{AB}$s are also normalized. Here $|\psi\rangle_{AB}(|\psi\rangle_{BC})$ is the state shared between Alice (Charlie) and Bob.}  {Using the vectors of Eq. (\ref{M2m}), we can write }	
 {\ba\label{J2qm}
J_{m,i}&=& \omega_{m,i}^{2}\langle M^{2}_{m,i}\rangle
\ea}
 { Here $|\psi\rangle_{AB}(|\psi\rangle_{BC})$ is the state shared between Alice (Charlie) and Bob. Also $\omega^{2}_{m,i}$ is defined as    $\omega^{2}_{m,i}=(\omega^{2}_{m,i})_{A}(\omega^{2}_{m,i})_{C}$.  Since $(\omega_{m,i}^{2})$s  are products of norms, $(\omega^{2}_{m,i})\geq 0, \forall i\in[m] $. Hence from Eq. (\ref{J2qm}), we can write}
 {\ba\na
(\Delta^{2}_{m})_{Q}&=&\sum\limits_{i=1}^m\sqrt{\omega_{m,i}^{2}|\langle M^{2}_{m,i}\rangle|}
\ea}

 { As we have defined, the vectors $M^{2}_{m,i}|\psi\rangle$  are normalized for each $i\in[m]$. Hence   the optimal value of $(\Delta^{2}_{m})_{Q}$ is obtained when  $\langle M^{2}_{m,i}\rangle=\pm 1$ for each $i\in[m]$. This ensures that   the state $|\psi\rangle$ has to be a pure state and it  is the eigenvector of each of the   observables  $M^{2}_{m,i}$ corresponding to the eigenvalue $\pm1$ 
 i.e.,  $M^{2}_{m,i}|\psi\rangle=\pm |\psi\rangle, \forall i\in[m]$. 
This implies that }
	\begin{eqnarray}(\Delta^{2}_{m})^{opt}_{Q}=\max_{A_x,C_z}\bigg(\sum\limits_{i=1}^m \sqrt{\omega^{2}_{m,i}}\bigg)\end{eqnarray}
 {The norms $(\omega^{2}_{m,i})_{A}$ and $(\omega^{2}_{m,i})_{C}$ are  given by
$(\omega^{2}_{m,i})_{A}=||\sum\limits_{x=1}^{2^{m}-1}(-1)^{y^{x}_{i}} A_{x}\hspace{1pt})|\psi\rangle_{AB}||_{2}$ and $(\omega^{2}_{m,i})_{C}=||(\sum\limits_{z=1}^{2^{m}-1}(-1)^{y^{z}_{i}} C_{z}\hspace{1pt})|\psi\rangle_{BC}||_{2}$.}

	  {Since  $\omega^{2}_{3,i}=(\omega^{2}_{3,i})_{A}(\omega^{2}_{3,i})_{C}, \forall i\in[m]$, by using the  inequality  Eq. (\ref{ATavakoli}),
	we get that}
 \ba \label{Awnmi}\sum\limits_{i=1}^m\sqrt{\omega^{2}_{m,i}}&\leq&\sqrt{\sum\limits_{i=1}^m(\omega^{2}_{m,i})_A}\sqrt{\sum\limits_{i=1}^m(\omega^{2}_{m,i})_C}\ea
	Further using the convex inequality, we have  \ba\label{Aconvexm}\sum\limits_{i=1}^m(\omega^{2}_{m,i})_A\leq \sqrt{m\sum\limits_{i=1}^m\bigg[(\omega^{2}_{m,i})_A\bigg]^2}\ea 
We have  obtained that \ba\label{optwAC}\max\limits_{A_x,x}\bigg(\sum\limits_{i=1}^{m}(\omega^{2}_{m,i})_{A}\bigg)= \max_{C_z,z
}\bigg(\sum\limits_{i=1}^{m}(\omega^{2}_{m,i})_{C}\bigg)=2^{m-1}\sqrt{m}\ea which implies that $(\Delta^{2}_{m})_{Q}^{opt}=2^{m-1}\sqrt{m}$.
 {Also we check that for the optimal value,  
\ba (\omega^{2}_{m,i})_{A}=(\omega^{2}_{m,i})_{C}=\frac{2^{m-1}}{\sqrt{m}}, \quad i\in[m]\ea} 

 The detailed derivation of this optimal bound in provided in Appendix \ref{B}.

 {Considering the optimal scenario,  let us denote}
 {\ba\mathcal{A}^{2,m}_i=\bigg(\frac{\sqrt{m}}{2^{m-1}}\sum\limits_{x=1}^{2^{m-1}}(-1)^{y^{x}_{i}} A_{x}\bigg)\otimes   \bigg(\frac{\sqrt{m}}{2^{m-1}}\sum\limits_{z=1}^{2^{m-1}}(-1)^{y^{z}_{i}} 
C_{z}\bigg)\ea}
 {Since the optimal quantum violation certifies a certain number of relations of the observables of Alice and Charlie, using  the relations Eq. (\ref{ajj'}) and Eq. (\ref{cjj'}), we obtain that}  {\ba
\label{aa}\mathcal{A}^{2,m}_i\mathcal{A}^{2,m }_j=\mathcal{A}^{2,m}_j\mathcal{A}^{2,m }_i, \quad \forall i\neq j\in[m]\ea}  {We can then write $M^{2}_{m,i}=\mathcal{A}_{i}^{2,m}\otimes B_i$.
 Since  $M^{2}_{m,i}|\psi\rangle=\pm|\psi\rangle, \forall i\in[m]$, the observables $M^{2}_{m,i}$, $M^{2}_{m,j}$ are mutually  commuting i.e.,   $\left[M^{2}_{m,i},M^{2}_{m,j}\right]=0, \forall i\neq j\in[m]$. 
Using Eq. (\ref{aa}), we get}
 {\ba\mathcal{A}^{2,m}_i\mathcal{A}^{2,m}_j\otimes \left(B_iB_j-B_jB_i\right)&=&0\ea}

That is, Bob's observables have to be mutually commuting to obtain the optimal quantum value. Since $m$ is arbitrary, the optimal quantum violation of bilocality inequality self-tests a set of an unbounded number of mutually commuting  {local} observables. 

However, the number of mutually commuting  observables is restricted by the dimension. For example, for a two-qubit system, at most three observables can be mutually commuting.  We find that to obtain the optimal value for arbitrary $m$, the local dimension of every Alice (Charlie) is to be at least $d=2^{\lfloor m/2\rfloor}$. In other words,   Alice (Charlie) shares at least $\lfloor m/2\rfloor$ copies of a two-qubit maximally entangled state with Bob. The total state  can be written as 
\begin{equation}
\label{state}
|\psi_{ABC}\rangle = |\phi^{+}_{AB}\rangle^{\otimes\lfloor{\frac{m}{2}\rfloor}}\otimes|\phi^{+}_{BC}\rangle^{\otimes\lfloor{\frac{m}{2}\rfloor}} 
\end{equation}
  
Thus, the optimal quantum violation of arbitrary input 
 bilocality inequality (\ref{deltam}) uniquely fixes Bob's observables of and  eventually  a set of an unbounded number of mutually commuting  {local} observables is self-tested.

\section{Summary and Discussion}
\label{V}
In summary, we provided schemes for self-testing an unbounded number of mutually commuting  {local} observables. It is a common perception that commuting observables cannot lead to nonclassicality as they have common eigenstates and  hence joint probability exists. It is also well-known that the demonstration of Bell's theorem requires incompatible observables. Since  commuting observables are compatible, and hence they cannot reveal quantum nonlocality. Against this backdrop, in this work, we demonstrated that the optimal quantum violation of network inequalities can only be obtained when the observables of one party are mutually commuting. Therefore, we showed that   the optimal quantum violation self-tests a set of mutually commuting  {local} observables. 

To demonstrate it, we first considered a star network involving an arbitrary $n$ number of edge parties and a central party. Each party including the central party receives two inputs $m=2$, and performs the measurement of two dichotomic observables accordingly. We showed that the optimal quantum violation of $n$-locality    inequality can only be obtained when the two observables of Bob are mutually commuting for $n$ is even. Importantly, we invoked an elegant  approach that enables us to derive the optimal quantum violation without any reference to the dimension of the system. In other words, the dimension of Bob's commuting observables remains unspecified. 

Further, we demonstrated the self-testing of an unbounded number of mutually commuting  {local} observables. This feature is also generic and valid for any arbitrary dimensional system. To demonstrate this, we considered a bilocal scenario $n=2$ where the central party Bob performs an arbitrary $m$ number of dichotomic measurements and each of the two edge parties performs  $2^{m-1}$ number of dichotomic measurements. We showed that optimal quantum violation of a suitably formulated bilocality inequality can be obtained only when Bob's observables are mutually commuting. Therefore, the optimal quantum value self-tests an unbounded number of mutually commuting  {local} observables as $m$ is arbitrary. 

The significance of this work is that it challenges the usual perception of commuting observables in quantum theory. Optimal quantum violation of Bell's inequality commonly certifies the anticommuting observables. Based on the optimal quantum violation of suitably formulated network inequalities, we demonstrated the self-testing of an unbounded number of mutually commuting  {local} observables. 

 \section{Acknowledgment}SM acknowledges the support from the research grant DST/ICPS/QuEST/2019/4. AKP acknowledges the support from the research grant  SERB/CRG/2021/004258.
\appendix \begin{widetext}
 \section{Deriavtion of the bilocal  bound of $(\eta^A_m)_{\max}$ in Eq. (\ref{etamA})}   \label{A}
Let us consider an expression $\mathcal{B}=\sum\limits_{i=1}^{m}\sum\limits_{x=1}^{2^{m-1}}(-1)^{y^{x}_{i}} A_{x}B_{i}$ where $A_x$ and $B_i$ are all dichotomic and the encoding scheme $y^{x}_{i}$ is same as depicted in main text Sec \ref{III}. Since each of $A_{x},B_i\in\{-1,1\}$, the observables $A_x$ and $B_i$ are basically equivalent,  the functional $\mathcal{B}$ is invariant under the interchange of indices $x$ and $i$. We can then  write $\mathcal{B}=\sum\limits_{i=1}^{2^{m-1}}\sum\limits_{x=1}^{m}(-1)^{y_{x}^{i}} A_{x}B_{i}$. 
Hence, using the fact that $|B_i|\leq 1, \forall i$, we get

\ba|\mathcal{B}|\leq\sum\limits_{i=1}^{m}\bigg|\sum\limits_{x=1}^{2^{m-1}}(-1)^{y^{x}_{i}} A_{x}\bigg|; \ \ |\mathcal{B}|\leq\sum\limits_{i=1}^{2^{m-1}}\bigg|\sum\limits_{x=1}^{m}(-1)^{y_{x}^{i}} A_{x}\bigg|\ea

Considering the tightness of the bound and the uniqueness of the  supremum property, we can write  \ba\sum\limits_{i=1}^{m}\bigg|\sum\limits_{x=1}^{2^{m-1}}(-1)^{y^{x}_{i}} A_{x}\bigg|= \sum\limits_{i=1}^{2^{m-1}}\bigg|\sum\limits_{x=1}^{m}(-1)^{y_{x}^{i}} A_{x}\bigg|\ea

	   Using the augmented Hadamard code \cite{had}, it was already derived in \cite{Sneha2021,snehachsh} that $\bigg(\sum\limits_{i=1}^{2^{m-1}}\bigg|\sum\limits_{x=1}^{m}(-1)^{y_{x}^{i}} A_{x}\bigg|\bigg)_{max}=m\binom{m-1}{\lfloor\frac{m-1}{2}\rfloor}$ which implies that $\bigg(\sum\limits_{i=1}^{m}\bigg|\sum\limits_{x=1}^{2^{m-1}}(-1)^{y^{x}_{i}} A_{x}\bigg)_{max}=m\binom{m-1}{\lfloor\frac{m-1}{2}\rfloor}$.  Since we considered $\sum\limits_{i=1}^{m}\bigg|\sum\limits_{x=1}^{2^{m-1}}(-1)^{y^{x}_{i}} A_{x}\bigg|=\eta^A_m$, clearly
    \ba(\eta^A_m)_{max}=m\binom{m-1}{\lfloor\frac{m-1}{2}\rfloor}\ea

\section{Detailed  derivation of the optimal quantum bound of $\sum\limits_{i=1}^{m}(\omega^{2}_{m,i})_{A}\bigg(\sum\limits_{i=1}^{m}(\omega^{2}_{m,i})_{C}\bigg)$ in Eq. (\ref{optwAC})}
\label{B}Since from Eq. (\ref{Awnmi}), we can write 
\begin{align} 
	\label{optbnmA}
	(\Delta^{2}_{m})_{Q}^{opt} &\leq \sqrt{ \bigg(\sum\limits_{i=1}^{m}(\omega^{2}_{m,i})_{A}\bigg)}\sqrt{ \bigg(\sum\limits_{i=1}^{m}(\omega^{2}_{m,i})_{C}\bigg)}\end{align} here we  optimize $\bigg(\sum\limits_{i=1}^{m}(\omega^{2}_{m,i})_{A}\bigg)$. Using the  convex inequality we can  write
\ba\sum\limits_{i=1}^{m}(\omega^{2}_{m,i})_{A}\leq\sqrt{m\sum\limits_{i=1}^{m}\bigg((\omega^{2}_{m,i})_{A}\bigg)^2}\ea
Using the definition of $(\omega^{2}_{m,i})_{A}$, we can write   $\sum\limits_{i=1}^{m}\bigg((\omega^{2}_{m,i})_{A}\bigg)^2=\langle\psi|m2^{m-1}+\delta_m|\psi\rangle$ where
\ba\label{delm}
\delta_m&=&\sum\limits_{l=1}^{2^{m-1}-m}(\delta_m)_l\\\nonumber
&=&(m-2)\sum\limits_{j'=2}^{1+\binom{m}{1}}\{A_1,A_{j'}\}+(m-4)\sum\limits_{j'=2+\binom{m}{1}}^{1+\binom{m}{1}+\binom{m}{2}}\{A_1,A_{j'}\}+\cdots+\bigg(m-2\lfloor\frac{m}{2}\rfloor\bigg)\sum\limits_{j'=2+\binom{m}{1}+\binom{m}{2}+\cdots\binom{m}{\lfloor\frac{m}{2}\rfloor-1} }^{1+\binom{m}{1}+\binom{m}{2}+\cdots\binom{m}{\lfloor\frac{m}{2}\rfloor}}\{A_1,A_{j'}\}+(m-4)\\&&\sum\limits_{j,j'=2 j\neq j'}^{1+\binom{m}{1}}\{A_j,A_{j'}\}
+\cdots\cdots +(m-4)
\{A_{2^{m-1}-1},A_{2^{m-1}} \}\hspace{18pt}\ea 
such that $(\delta_m)_l=2^{m-1}-\langle\psi_l|\psi_l\rangle$. Hence, 
\ba \delta_m=(2^{m-1}-m)2^{m-1}-\sum\limits_{l=1}^{2^{m-1}-m}\langle\psi_l|\psi_l\rangle\ea
 where we define 
  \ba\label{ll}|\psi_l\rangle=\sum\limits_{x=1}^{2^{m-1}} (-1)^{s_l.y^{x}}A_{x}|\psi\rangle.\ea 
 
 To find the element $s_l$, we  consider a set $\mathcal{L}_m=\{s|s\in\{0,1\}^m, \sum_r s_r\geq 2\}$, $r\in\{1,2,\cdots m\}$. The element  $s_l\in \mathcal{L}_m$ is  such that $\sum_r(s_l)_r\neq 2u$, for some $u\in\mathbb{N}$. We then find $(2^{m-1}-m)$ number of $s_l$  where $l\in[2^{m-1}-m]$.  Clearly, 	$(\delta_m)_{max}=(2^{m-1}-m)2^{m-1}$ and it holds only when for each $l\in[2^{m-1}-m]$, $|\psi_l\rangle=0$. Since $|\psi\rangle\neq 0$, hence for optimization, the observables for each Alice must satisfy the conditions  $\sum\limits_{x=1}^{2^{m-1}} (-1)^{s_l.y^{x}}A_{x}=0$, for each $l\in[2^{m-1}-m]$.
  Finally, we get  $\sum\limits_{i=1}^{m}\bigg((\omega^{2}_{m,i})_{A}\bigg)_{opt}^2=2^{2(m-1)}$ which in turn provides $\sum\limits_{i=1}^{m}\bigg((\omega^{2}_{m,i})_{A}\bigg)_{opt}=2^{m-1}\sqrt{m}$.
   The observables of each Alice satisfy the condition
	\ba\label{ajj'}\{A_j,A_{j'}\}=2 - \frac{4p}{m}, \forall j,j'=x\in[2^{m-1}] \ea     Clearly, there exists the $j(j')^{th}$ bit string denoted by $y^{j}(y^{{j'}})$ from the set of $2^{m-1}$ bit strings as defined earlier. Let the set  $\{y^{j}\}$ contains all the elements ($0$ or $1$) of that corresponding bit string. 
	Hence, for $x=j(j')\in[2^{m-1}]$, we can consider the set $\{y^{j}\}\cup \{y^{{j'}}\}$ as the collection of those elements corresponds to the bit strings   $\{y^{j}\}$ and $\{y^{j'}\}$. Without loss of generality, let us assume, $\{y^{j}\}\cup \{y^{{j'}}\}$ contains  $q$ number of $1$ s in it. Clearly, from the construction of the bit strings, here $0\leq q\leq m$.  
	
	Now we  divide the bit strings $y^{j}(y^{j'})$ into $(\lfloor\frac{m}{2}\rfloor+1)$ classes according to the number of $1$s in it. Let $y^{j}\in C^{\nu}$ if the corresponding bit string of $y^{j}$ contains $\nu$ number of $1$s in it.  Let there are two classes $C^{\nu}$ and $C^{\nu'}$ such that $y^{j}\in C^{\nu}$ and $y^{j'}\in C^{\nu'}$ and $\nu+\nu'=q$ ($0\leq \nu, \nu'\leq\lfloor\frac{m}{2}\rfloor$). For a given pair of $(j,j')$, there exists $t\in \mathbb{T}\subseteq [m] $ such that $y^j_t=y^{j'}_t=1$. Let the cardinality of the set $\mathbb{T}$ i.e.,  $|\mathbb{T}|=d$. Then there exists a number $p$ such that  $p=q-2d$. Using it in Eq. (\ref{ajj'}), we get the observables for each Alice$_k$.
Similarly, we can find 
\ba\label{cjj'}\{C_j,C_{j'}\}=2-\frac{4p}{m}, \forall j(j')=z\in[2^{m-1}] \ea 
For example, we can consider the $m=3$ scenario. Considering Eq. (\ref{delm}), here we can write 
\ba \delta_3=\{A_{1},(A_{2}+A_{3}+A_{4})\}-\{A_{2},(A_{3}+A_{4})\}-\{A_{3},A_{4}\})\ea
as derived in Eq. (\ref{del3main}).  Here the set $\mathcal{L}_3$ is defined as  $\mathcal{L}_3=\{s|s\in\{0,1\}^3, \sum_r s_r\geq 2\}$, $r\in\{1,2,3\}$. Hence here $\mathcal{L}_3$ contains the elements $\{110,101,011,111\}$ only. We denote the  element  $s_l\in \mathcal{L}_3$  such that $\sum_r(s_l)_r\neq 2u$, for some $u\in\mathbb{N}$. Hence here we then find $(2^{3-1}-3)=1$ element  $s_1$ which is $s_1=111$.
. Clearly, 	$(\delta_3)_{max}=(2^{3-1}-3)2^{3-1}=4$ and it holds only when  $|\psi_1\rangle=\sum\limits_{x=1}^{4} (-1)^{s_1.y^{x}}A_{x}|\psi\rangle=0$. Since $|\psi\rangle\neq 0$, for optimization, the observables for each Alice must satisfy the conditions  $\sum\limits_{x=1}^{4} (-1)^{s_1.y^{x}}A_{x}=0$. Since  here $s_1=111$, 
$\sum\limits_{x=1}^{4} (-1)^{s_1.y^{x}}A_{x}=0\implies A_1-A_2-A_3-A_4=0$ which is already derived in Sec \ref{III}.
  Finally, we get  $\sum\limits_{i=1}^{3}\bigg((\omega^{2}_{3,i})_{A}\bigg)_{opt}^2=2^{2(3-1)}=16$ which in turn provides $(\Delta^{2}_{3})_{Q}^{opt}=4\sqrt{3}$.\\
  The observables of each Alice satisfy the condition  
	\ba 
 \label{ajj'3}
 \{A_j,A_{j'}\}=2-\frac{4p}{3}\ea  
 where $j,j'=x\in[4]$. Clearly, there exists the $j(j')^{th}$ bit string  denoted by $y^{j}(y^{{j'}})$ from the set of $4$ bit strings as defined earlier. Let the set  $\{y^{j}\}$ contains all the elements ($0$ or $1$) of that corresponding bit string.
 Let us consider that $j=1,j'=2$. Hence the corresponding $3$ length bit strings are $y^1=000$ and $y^2=001$. Hence, we can consider the set $\{y^{1}\}\cup \{y^{{2}}\}$ as the collection of those elements corresponding to the bit strings   $y^{1}$ and $y^{2}$. Clearly we can see that  $\{y^{1}\}\cup \{y^{{2}}\}=\{0,0,0,0,0,1\}$ contains  a single  $1$ in it. Hence, in this case $q=1$. 
	
	Since  we  divide the bit strings $y^{j}(y^{j'})$ into $(\lfloor\frac{3}{2}\rfloor+1)=2$ classes according to the number of $1$s in it, here we can see that  $y^{1}\in C^{\nu=0}$ and  $y^{2}\in C^{\nu'=1}$ and here $\nu+\nu'=q=1$. For a given pair of $(1,2)$, there exists $t\in \mathbb{T}\subseteq [3] $ such that $y^1_t=y^{2}_t=1$. Since $y^1=000$ and $y^2=001$, there is no such $t$ such that $y^1_t=y^{2}_t=1$. Hence  the cardinality of the set $\mathbb{T}$ is $0$ i.e.,  $|\mathbb{T}|=d=0$. Then we get $p=q-2d=1$. Using it in Eq. (\ref{ajj'3}), we get the
 \ba 
 \label{ajj'3new}
 \{A_1,A_{2}\}=2-\frac{4}{3}=\frac{2}{3}\ea
as derived in Sec \ref{III}. Similarly, we can find all the anti-commutation relations of both Alice and Charlie.

  \end{widetext}


\begin{thebibliography}{99}
\bibitem{buschrep}P. Busch, T. Heinonen, P. Lahti, Heisenberg's Uncertainty Principle, \href{
https://doi.org/10.1016/j.physrep.2007.05.006
}{ Phys. Rep., \textbf{452}, 155 (2007)}.
\bibitem{bell} J.S. Bell, On the Einstein Podolsky Rosen paradox, \href{https://doi.org/10.1103/PhysicsPhysiqueFizika.1.195}{Physics, {\bf 1}, 195 (1964)}.
\bibitem{chsh} J. F. Clauser, M. A. Horne, A. Shimony, R. A. Holt, Proposed Experiment to Test Local Hidden-Variable Theories, 
		\href{http://doi.org/10.1103/PhysRevLett.23.880}{ Phys. Rev. Lett. {\bf 23}, 880 (1969)}.

\bibitem{wolf} M. M. Wolf, D. Perez-Garcia and C. Fernandez, Measurements Incompatible in Quantum Theory Cannot Be Measured Jointly in Any Other No-Signaling Theory,\href{https://journals.aps.org/prl/abstract/10.1103/PhysRevLett.103.230402}{ Phys. Rev. Lett. \textbf{103}, 230402 (2009)}.
\bibitem{bene} E. Bene and T. Vertesi, Measurement incompatibility does not give rise to Bell violation in general, \href{https://iopscience.iop.org/article/10.1088/1367-2630/aa9ca3/meta}{  New J. Phys. \textbf{20}, 013021 (2018)}.

			
			\bibitem{hororev} R. Horodecki, P . Horodecki, M. Horodecki, and K. Horodecki, Quantum entanglement, \href{https://doi.org/10.1103/RevModPhys.81.865} {Rev. Mod. Phys. \textbf{81}, 865 (2009)}.
			\bibitem{guhnearev} O. G\"{u}hne and G. T\'{o}th, Entanglement detection, \href{https://doi.org/10.1016/j.physrep.2009.02.004} {Phys. Rep. \textbf{474}, 1 (2009)}.
			\bibitem{brunnerrev} N. Brunner, D. Cavalcanti, S. Pironio, V. Scarani, and S. Wehner, Bell nonlocality, \href{https://doi.org/10.1103/RevModPhys.86.419}{Rev. Mod. Phys. \textbf{86}, 419 (2014)}.
			
			
			\bibitem{bar05} J. Barrett, L. Hardy and A. Kent, No Signaling and Quantum Key Distribution, \href{https://journals.aps.org/prl/abstract/10.1103/PhysRevLett.95.010503}{Phys. Rev. Lett. \textbf{95}, 010503(2005).}
			\bibitem{acin06}A. Acin, N. Gisin and L. Masanes, From Bell’s Theorem to
			Secure Quantum Key Distribution, \href{https://journals.aps.org/prl/abstract/10.1103/PhysRevLett.97.120405}{Phys. Rev. Lett. \textbf{97}, 120405 (2006).}
			\bibitem{acin07} A. Acin, N. Brunner, N. Gisin, S. Massar, S. Pironio and and V.
			Scarani, Device-Independent Security of Quantum Cryptography against Collective Attacks, \href{https://journals.aps.org/prl/abstract/10.1103/PhysRevLett.98.230501}{ Phys. Rev. Lett. \textbf{98}, 230501 (2007).} 
			\bibitem{pir09}S. Pironio, A. Acin, N. Brunner, N. Gisin, S. Massar and V.
			Scarani, Device-independent quantum key distribution secure against collective attacks, \href{https://iopscience.iop.org/article/10.1088/1367-2630/11/4/045021/meta}{New J. Phys. \textbf{11}, 045021 (2009).} 
			
			\bibitem{col06} R. Colbeck, Quantum and relativistic protocols for secure multi-party computation, Ph.D. thesis, University of Cambridge (2006); \href{https://arxiv.org/abs/0911.3814}{arXiv:0911.3814v2.} 
			\bibitem{pir10} S. Pironio, et al., Random numbers certified by Bell’s theorem,\href{https://www.nature.com/articles/nature09008}{ Nature volume \textbf{464},1021(2010).}
			\bibitem{nieto} O. Nieto-Silleras, S. Pironio and J. Silman, Using complete measurement statistics for optimal device-independent randomness evaluation, \href{https://iopscience.iop.org/article/10.1088/1367-2630/16/1/013035}{New J. Phys. \textbf{16}, 013035 (2014).} 
			\bibitem{col12}R. Colbeck and R. Renner, Free randomness can be amplified, \href{https://www.nature.com/articles/nphys2300}{ Nature Physics \textbf{8}, 450(2012).
			}



   
			\bibitem{wehner} S. Wehner, M. Christandl and A.C. Doherty, Lower bound on the dimension of a quantum system given measured data, \href{https://journals.aps.org/pra/abstract/10.1103/PhysRevA.78.062112}{Phys. Rev. A \textbf{78},062112 (2008).} 
			\bibitem{gallego} R. Gallego, N. Brunner, C. Hadley, and A. Acin, Device-Independent Tests of Classical and Quantum Dimensions, \href{https://journals.aps.org/prl/abstract/10.1103/PhysRevLett.105.230501}{Phys. Rev. Lett. \textbf{105}, 230501 (2010).} 
			\bibitem{ahrens} J. Ahrens, P. Badziag, A. Cabello and M. Bourennane, Experimental device-independent tests of classical and quantum dimensions, \href{https://www.nature.com/articles/nphys2333}{Nat.Phys, \textbf{8}, 592(2012).} 
			\bibitem{brunnerprl13} N. Brunner, M. Navascues and T. Vertesi, Dimension Witnesses and Quantum State Discrimination, \href{https://journals.aps.org/prl/abstract/10.1103/PhysRevLett.110.150501}{Phys. Rev. Lett. \textbf{110}, 150501 (2013).} 
			
			\bibitem{bowler} J. Bowles, M. Quintino, and N. Brunner Certifying the Dimension of Classical and Quantum Systems in a Prepare-and-Measure Scenario with Independent Devices, +
   \href{https://journals.aps.org/prl/abstract/10.1103/PhysRevLett.112.140407}{Phys. Rev. Lett. \textbf{112}, 140407 (2014).
			}
			\bibitem{sik16prl}J. Sikora, A. Varvitsiotis and Z. Wei, Minimum Dimension of a Hilbert Space Needed to Generate a Quantum Correlation, \href{https://journals.aps.org/prl/abstract/10.1103/PhysRevLett.117.060401}{Phys. Rev. Lett., \textbf{117}, 060401 (2016)} 
			
			\bibitem{cong17} W. Cong, Y. Cai, J-D. Bancal and V. Scarani, Witnessing Irreducible Dimension, \href{https://journals.aps.org/prl/abstract/10.1103/PhysRevLett.119.080401}{Phys. Rev. Lett. 119, 080401 (2017).} 
			
			\bibitem{pan2020} A. K. Pan and S. S. Mahato, Device-independent certification of the Hilbert-space dimension using a family of Bell expressions,\href{https://doi.org/10.1103/PhysRevA.102.052221}{Phys. Rev. A \textbf{102}, 052221(2020)}.
			
			\bibitem{complx1}  H. Buhrman, R. Cleve, S. Massar,  and R. de Wolf, Nonlocality and communication complexity, \href{https://journals.aps.org/rmp/pdf/10.1103/RevModPhys.82.665}{Phys. Rev. Lett. \textbf{114}, 250401 (2015)}
			

   \bibitem{mayer98} D. Mayers, and A. Yao, Quantum cryptography with imperfect apparatus, in Proceedings of the 39th IEEE Conference on Foundations of Computer Science,\href{https://ieeexplore.ieee.org/document/743501}{ Palo Alto, CA, 1998 (IEEE, New York, 1998)}.

   \bibitem{supicrev} I. Supic and J. Bowles, Self-testing of quantum systems: A review,\href{
https://doi.org/10.22331/q-2020-09-30-337
} { Quantum,\textbf{4}, 337 (2020)}.
			
			
			
   


\bibitem{mckague}M. McKague and M. Mosca, Generalized self-testing and the security of the 6-state protocol,\href{
https://doi.org/10.1007/978-3-642-18073-6_10
}{Theory of Quantum Computation, Communication, and Cryptography, 2011, Volume 6519, ISBN : 978-3-642-18072-9(2010)}.
\bibitem{mayers} D. Mayers and A. Yao, Self testing quantum apparatus,  \href{https://dl.acm.org/doi/10.5555/2011827.2011830}{Quan. Inf. Comp. \textbf{4}, 273 (2004)}.


\bibitem{wagner2020} S. Wagner, J.D. Bancal, N. Sangouard, and P. Sekatski, \href{https://quantum-journal.org/papers/q-2020-03-19-243/pdf/}{Quantum \textbf{4}, 243 (2020).}
   \bibitem{farkas2019} M. Farkas and J. Kaniewski, Self-testing mutually unbiased bases in the prepare-and-measure scenario, \href{https://journals.aps.org/pra/abstract/10.1103/PhysRevA.99.032316}{Phys. Rev. A \textbf{99}, 032316 (2019).}
\bibitem{mckague12}M. McKague, T. H. Yang, and V. Scarani, Robust Self Testing of the Singlet,   \href{
https://doi.org/10.1088/1751-8113/45/45/455304
}{J. Phys. A: Math. Theor. \textbf{45}, 455304 (2012)}.

\bibitem{wu} X. Wu, J.D. Bancal, M. McKague, V. Scarani, Device-independent parallel self-testing of two singlets, \href{https://journals.aps.org/pra/abstract/10.1103/PhysRevA.93.062121}{Phys. Rev. A, \textbf{93}, 062121 (2016)}.
\bibitem{mckague16} M. McKague, Self-testing in parallel, \href{https://iopscience.iop.org/article/10.1088/1367-2630/18/4/045013}{New J. Phys. \textbf{18} 045013(2016)}.

\bibitem{and17}O. Andersson, P. Badziag, I. Bengtsson, I. Dumitru, and A. Cabello, Self-testing properties of Gisins elegant Bell inequality,
\href{https://journals.aps.org/pra/abstract/10.1103/PhysRevA.96.032119}{Phys. Rev. A \textbf{96}, 032119(2017)}.


\bibitem{cola}A. Coladangelo, K. T. Goh, and V. Scarani, All pure bipartite entangled states can be self-tested, \href{https://www.nature.com/articles/ncomms15485}{Nat. Commun. \textbf{8}, 15485 (2017)}.


\bibitem{supic18} I. Supic, A. Coladangelo, R. Augusiak and  A. Acin, Self-testing multipartite entangled states through projections onto two systems, \href{https://iopscience.iop.org/article/10.1088/1367-2630/aad89b/meta}{ New J. Phys. \textbf{20} 083041 (2018)}.
\bibitem{bowels18prl}J.Bowles, I.Supic, D. Cavalcanti and A. Acin, Device-Independent Entanglement Certification of All Entangled States, \href{https://doi.org/10.1103/PhysRevA.121.180503}{Phys. Rev. Lett. \textbf{121}, 180503 (2018)}. 
\bibitem{bowles18} J.Bowles, I.Supic, D. Cavalcanti and A. Acin, Self-testing of Pauli observables for device-independent entanglement certification,\href{
https://doi.org/10.1103/PhysRevA.98.042336
}{ Phys. Rev. A \textbf{98}, 042336 (2018)}.
\bibitem{coopmans} T. Coopmans, J. Kaniewski, and C. Schaffner, Robust self-testing of two-qubit states,\href{https://journals.aps.org/pra/abstract/10.1103/PhysRevA.99.052123} {Phys. Rev. A \textbf{99}052123 (2019)}.
\bibitem{tavakoli19a} A.Tavakoli, M. Farkas, D. Rosset, J-D Bancal, and J. Kaniewski, Mutually unbiased bases and symmetric informationally complete measurements in Bell experiments: Bell inequalities, device-independent certification and applications,\href{
https://doi.org/10.1126/sciadv.abc3847} {Science Advances \textbf{7}, eabc3847 (2021)}.




\bibitem{tavakoli18}  A. Tavakoli, J. Kaniewski, T. Vertesi, D. Rosset, and N. Brunner, Self-testing quantum states and measurements in the prepare-and-measure scenario, \href{https://journals.aps.org/pra/abstract/10.1103/PhysRevA.98.062307}{Phys. Rev. A \textbf{98}, 062307 (2018)}.
\bibitem{farkas} M. Farkas and J. Kaniewski, Self-testing mutually unbiased bases in the prepare-and-measure scenario, \href{https://journals.aps.org/pra/abstract/10.1103/PhysRevA.99.032316}{Phys. Rev. A \textbf{99}, 032316 (2019)}.



\bibitem{mir19}P. Mironowicz and M. Pawlowski, Experimentally feasible semi-device-independent certification of four-outcome positive-operator-valued measurements,\href{https://journals.aps.org/pra/abstract/10.1103/PhysRevA.100.030301}{ Phys. Rev. A \textbf{100}, 030301 (2019)}.
\bibitem{tavakoli20} A.Tavakoli, M. Smania, T. Vertesi, N. Brunner, M.Bourennane, Self-testing non-projective quantum measurements in prepare-and-measure experiments, \href{https://www.science.org/doi/10.1126/sciadv.aaw6664}{Sci. Adv. \textbf{6}, 16 (2020)}.
\bibitem{smania} M. Smania, P. Mironowicz, M. Nawareg, M. Pawlowski, A. Cabello, M. Bourennane, Experimental certification of an informationally complete quantum measurement in a device-independent protocol, \href{https://opg.optica.org/optica/fulltext.cfm?uri=optica-7-2-123&id=426348}{Optica, \textbf{7}, 123 (2020)}.
\bibitem{paw20} N. Miklin, J. J. Borkala, and M. Pawlowski, Semi-device-independent self-testing of unsharp measurements,\href{https://journals.aps.org/prresearch/abstract/10.1103/PhysRevResearch.2.033014} {Phys. Rev. Research \textbf{2}, 033014 (2020)}.
\bibitem{gomez16} E. S. Gomez \emph{et al.,} Device-Independent Certification of a Nonprojective Qubit Measurement, \href{https://journals.aps.org/prl/abstract/10.1103/PhysRevLett.117.260401}{Phys. Rev. Lett., \textbf{117}, 260401 (2016)}.


  \bibitem{gomez18}S. Gomez, A. Mattar, E. S. Gomez, D. Cavalcanti, O. Jimenez Farias, A. Acin, and G. Lima,  Experimental nonlocality-based randomness generation with nonprojective measurements, \href{https://journals.aps.org/pra/abstract/10.1103/PhysRevA.97.040102}{Phys. Rev. A \textbf{97}, 040102(R) (2018)}.
  \bibitem{pan2021} A. K. Pan, Oblivious communication game, self-testing of projective and nonprojective measurements, and certification of randomness, \href{https://doi.org/10.1103/PhysRevA.104.022212}{Phys. Rev. A, \textbf{104},022212, (2021)}.
   \bibitem{kartik} K. Mohan, A. Tavakoli and N. Brunner, Sequential random access codes and self-testing of quantum measurement instruments,\href{https://iopscience.iop.org/article/10.1088/1367-2630/ab3773} {New J. Phys. \textbf{21} 083034 (2019)}.
 \bibitem{Miklin2020} N. Miklin, J. J. Borkała, and M. Pawłowski, Semi-deviceindependent self-testing of unsharp measurements, \href{https://journals.aps.org/prresearch/pdf/10.1103/PhysRevResearch.2.033014}{ Phys. Rev.
Research \textbf{2}, 033014 (2020)}.
\bibitem{Anwar2020} H. Anwer, S. Muhammad, W. Cherifi, N. Miklin, A. Tavakoli,
and M. Bourennane, Experimental Characterization of Unsharp
Qubit Observables and Sequential Measurement Incompatibility via Quantum RAC, \href{https://journals.aps.org/prl/abstract/10.1103/PhysRevLett.125.080403}{Phys. Rev. Lett. \textbf{125}, 080403 (2020)}.
\bibitem{Foletto2020} G. Foletto, L. Calderaro, G. Vallone, and P. Villoresi, Experimental demonstration of sequential quantum random access
codes, \href{https://journals.aps.org/prresearch/abstract/10.1103/PhysRevResearch.2.033205}{Phys. Rev. Research \textbf{2}, 033205 (2020)}.
	\bibitem{sumit21} S. Mukherjee and A. K. Pan, Semi-device-independent certification of multiple unsharpness parameters through sequential measurements, \href{https://journals.aps.org/pra/abstract/10.1103/PhysRevA.104.062214}{Phys. Rev. A \textbf{104}, 062214 (2021)}.
 \bibitem{Abhy2023}Abhyoudai S. S., S.  Mukherjee, and A. K. Pan, Robust certification of unsharp instruments through sequential quantum advantages in a prepare-measure communication game
\href{https://journals.aps.org/pra/abstract/10.1103/PhysRevA.107.012411}{
Phys. Rev. A \textbf{107}, 012411( 2023)}.
	\bibitem{Branciard10} ] C. Branciard, N. Gisin, and S. Pironio, Characterizing the Nonlocal Correlations Created via Entanglement Swapping, \href{https://journals.aps.org/prl/abstract/10.1103/PhysRevLett.104.170401}{Phys. Rev. Lett. {\bf 104},170401 (2010)}.
\bibitem{Branciard12} C. Branciard, D. Rosset, N. Gisin, and S. Pironio, Bilocal versus nonbilocal correlations in entanglement-swapping experiments, \href{https://journals.aps.org/pra/abstract/10.1103/PhysRevA.85.032119}{Phys. Rev. A {\bf 85}, 032119 (2012).}
	
\bibitem{prabuddha} P. Roy and A. K. Pan, Device-independent certification of unsharp measurements, \href{https://doi.org/10.1088/1367-2630/acb4b5}{ New J Phys. (2023)}.
 {\bibitem{Nava2020}
M. Navascués, E. Wolfe, D. Rosset, and A. P. Kerstjens, Genuine Network Multipartite Entanglement, \href{https://journals.aps.org/prl/abstract/10.1103/PhysRevLett.125.240505}{ Phys. Rev. Lett.\textbf{ 125}, 240505 (2020).}}
 {\bibitem{Agre2021}
I. Agresti \emph{et. al.}, Experimental Robust Self-Testing of the State Generated by a Quantum Network,  \href{https://journals.aps.org/prxquantum/abstract/10.1103/PRXQuantum.2.020346}{ PRX Quantum \textbf{2}, 020346 (2021).}}
 {\bibitem{Supic2021}I. Supic , J. D.  Bancal ,Y. Cai ,
 and N. Brunner, 
 Genuine network quantum nonlocality and self-testing,
		\href{https://doi.org/10.1103/PhysRevA.105.022206}{Phys.
			Rev. A {\bf 105}, 022206(2021)}.}	
 {\bibitem{Supic2023} I. Supić, J.  Bowles, M-O. Renou, A. Acín, and  M.  J. Hoban, Quantum networks self-test all entangled states, \href{https://doi.org/10.1038/s41567-023-01945-4}{ Nat. Phys. \textbf{8}, 450(2023).}}


  
  \bibitem{Armi2014} A. Tavakoli, P. Skrzypczyk, D. Cavalcanti, and A. Acin, Nonlocal correlations in the star-network configuration,   \href{https://doi.org/10.1103/PhysRevA.90.062109}  {Phys. Rev. A {\bf 90}, 062109 (2014)}.
  	
\bibitem{Ambainis} A. Ambainis, D. Leung, L. Mancinska, M. Ozols, Quantum Random Access Codes with Shared Randomness, 
		\href{https://arxiv.org/abs/0810.2937}{ arXiv:0810.2937v3}.
		\bibitem{Ghorai2018} S. Ghorai, A. K. Pan, Optimal quantum preparation contextuality in an 
$n$-bit parity-oblivious multiplexing task,
		\href{https://doi.org/10.1103/PhysRevA.98.032110}{Phys.
			Rev. A {\bf 98}, 032110(2018)}.			
   \bibitem{had} 
 A. Sanjeev, B. Boaz, Computational Complexity: A Modern Approach. Cambridge University Press, ISBN 978-0-521-42426-4.
 \bibitem{Sneha2021} S. Munshi, R. Kumar, A. K. Pan, Generalized 
$n$
-locality inequalities in a star-network configuration and their optimal quantum violations,  
\href{https://doi.org/10.1103/PhysRevA.104.042217}{ Phys. Rev. A {\bf 104}, 042217(2021)}.
\bibitem{snehachsh}S. Munshi and A. K. Pan, Characterizing nonlocal correlations through various n-locality inequalities in quantum network \href{https://journals.aps.org/pra/abstract/10.1103/PhysRevA.105.032216}{ Phys.
			Rev. A {\bf 105}, 032216 (2022)}.
 
\end{thebibliography}
\end{document}